\documentclass[%
superscriptaddress,
preprint,
showkeys,
 amsmath,amssymb,
aps,
floatfix
]{revtex4-1}

\usepackage{subfigure}
\usepackage{graphicx}
\usepackage{tikz}
\usepackage{dcolumn}
\usepackage{bm}
\usepackage{multirow}
\usepackage{booktabs}
\usepackage{physics}
\usepackage{tabularx}
\usepackage{tabulary}

\usepackage{cancel}

\usepackage{ulem}
\usepackage{gnuplottex}
\usepackage{mhchem}

\usepackage[]{xcolor}
\usepackage{xspace}
\newcolumntype{Y}{>{\centering\arraybackslash}X}

\newcommand*{\wfqfmu}{$\omega$FQF$\mu$\xspace}

\makeatletter
\def\@fnsymbol#1{%
  \ensuremath{%
    \ifcase#1\or
      \dagger\or
      *\or
      \ddagger\or
      \mathsection\or
      \mathparagraph\or
      \|\or
      **\or
      \dagger\dagger\or
      \ddagger\ddagger
    \else
      \@ctrerr
    \fi
  }%
}
\makeatother

\graphicspath{{}}

\begin{document}


\keywords{Chirality, Plasmonics, Quantum, Classical, Atomistic, Modeling, Gold, Silver, Nanoparticles\vspace{-3cm}
}

\title{Beyond the Quantum Picture: The Electrodynamic Origin of Chiral Nanoplasmonics}

\author{Vasil Saroka}
\thanks{V.S. and L.C. contributed equally to this work.}
\affiliation{Department of Physics, University of Rome Tor Vergata, Via della Ricerca Scientifica 1, 00133, Rome, Italy}
\author{Lorenzo Cupellini}
\thanks{V.S. and L.C. contributed equally to this work.}
\affiliation{Dipartimento di Chimica e Chimica Industriale, Università di Pisa, Via G. Moruzzi 13, Pisa, 56124, Italy}
\author{Nicolò Maccaferri}
\affiliation{Ultrafast Nanoscience Group, Department of Physics, Umeå University, Umeå, Sweden.}
\affiliation{Wallenberg Initiative Materials Science for Sustainability, Department of Physics, Umeå University, Umeå, Sweden.}
\author{Alessandro Fortunelli}
\affiliation{CNR Institute of Chemistry of OrganoMetallic Compounds (CNR-ICCOM), Via Giovanni Moruzzi 1, 56124, Pisa, Italy}
\author{Tommaso Giovannini}
\email{tommaso.giovannini@uniroma2.it}
\affiliation{Department of Physics, University of Rome Tor Vergata, Via della Ricerca Scientifica 1, 00133, Rome, Italy}

\begin{abstract}
Chiral plasmonic nanostructures are rapidly emerging as ideal substrates for enantioselective sensing, chiral near-field engineering, and plasmon-assisted catalysis, owing to their exceptional sensitivity to structural handedness. However, the physical origin of plasmonic chirality, whether intrinsically quantum or primarily governed by collective electrodynamics, remains an open question, limiting the development of predictive theoretical methods for the design of novel chiral plasmonic architectures. Here, we show that a fully atomistic classical electrodynamic model, coupling intraband charge transport and interband polarization, quantitatively reproduces state-of-the-art \textit{ab initio} and experimental chiroptical spectra across the quantum-to-classical regime, from atomistically defined chiral Ag and Au nanostructures to DNA-origami-assembled Au nanorods containing up to $\sim 10^5$ atoms. Our results support a unified electrodynamic origin of plasmonic chirality, providing the missing foundation to connect local structural motifs to chiroptical response and local chiral near fields, and paving the way for the atomistically defined, rational design of chiral plasmonic nanostructures optimized for targeted applications.
\end{abstract}

\maketitle


\section{Introduction}

Chiral plasmonic nanostructures are a promising platform to explore and exploit optical activity at the nanoscale.\cite{wu2022chiral,hentschel2017chiral,yin2015active,govorov2011chiral,ben2013chirality,valev2013chirality,neubrech2020reconfigurable,kuzyk2012dna,kim2019assembly,shen2013three} By combining strong light--matter coupling in metallic architectures with broken mirror symmetry, they can generate intense chiroptical signals, generally measured in terms of circular dichroism (CD),\cite{fan2012chiral,fan2010plasmonic,chow2026intrinsic,kwon2023chiral,wang2017circular,Yang2025,yao2025chiral,schreiber2013chiral,kuzyk2012dna} as well as circularly polarized near fields, enabling applications in chiral sensing, nanophotonics, metamaterials, and catalysis.\cite{solomon2020nanophotonic,both2022nanophotonic,serrano2026chiral,zhang2024self,bainova2023plasmon,wang2016optical} At a more fundamental level, chiral plasmonics offers a direct route to probe how collective plasmonic resonances reflect geometric asymmetry and thus to investigate the physical origin of optical chirality.\cite{avalos2022chiral} In fact, plasmonic excitations are collective and delocalized over the whole structure.\cite{odom2011introduction} However, atomic-scale defects can qualitatively alter the optical and chiroptical responses, which are indeed extremely sensitive to subtle variations in atomic-scale morphology,\cite{baumberg2022picocavities,giovannini2025electric,girod2025three,kong2020plasmonic,urbieta2018atomic} surface corrugation, interparticle distance, and relative orientation within a chiral assembly.\cite{vila2022template} 
Such an atomistic sensitivity is particularly relevant for hybrid plasmon-molecule systems, where the intensity, handedness, and spatial localization of the chiral near field at the metal interface can directly control molecular chiroptical enhancement, chirality transfer, and plasmon-assisted photochemical processes.

Generally, for atomistically defined nanostructures, an explicitly quantum-mechanical (QM) description is regarded as the most natural framework to connect structure and chiroptical spectra.\cite{monti2023contributes,toffoli2021circularly,toffoli2021plasmonic,krishnadas2020chiral,d2024dichroism,harkonen2024enhancement,makkonen2021real,santizo2008intrinsic,noguez2014ab,hodgins2024emergence} In this context, time-dependent density functional theory (TDDFT) provides a rigorous and accurate approach for small metallic nanoparticles and is currently considered the state-of-the-art tool for the prediction of (chiro)optical response. However, chiral plasmonic architectures of experimental interest typically involve thousands to tens of thousands of atoms, for which TDDFT is computationally prohibitive. In this size regime, the optical activity is generally interpreted by means of continuum electrodynamical methods,\cite{fan2010plasmonic,fan2012chiral,carone2022insight} where the response is described in terms of long-range electromagnetic coupling distributed over the entire nanostructure, and atomic resolution is typically neglected,\cite{Slepyan1999,Shuba2024} implicitly assuming that atomic-scale structure plays a minor role. Yet atomistic morphology (surface roughness, defects, and subnanometer gaps) can strongly modulate the collective plasmonic modes\cite{urbieta2018atomic,giovannini2025electric,baumberg2022picocavities,zheng2021discrete,sementa2014atomistic} and may be crucial to describe local chiral near fields, especially in hybrid plasmon--molecule systems. Reconciling these aspects requires a framework that can treat large nanostructures while incorporating the correct local physics, which, for atomistically defined systems, is generally accessible only to QM methods. This limitation becomes even more severe for hybrid plasmon--molecule complexes, and leaves a fundamental question unanswered: whether plasmonic chirality is intrinsically quantum in origin, or instead emerges from collective electrodynamics governed by atomic-scale morphology. Answering this question is essential not only for establishing the theoretical foundation of predictive models for realistic chiral plasmonic architectures, but also for identifying the minimal level of theory required to describe hybrid plasmon--molecule systems, where the nanoparticle response and the molecular quantum degrees of freedom must be treated on distinct, yet physically consistent, footings.

Here, we address this gap by challenging a fully atomistic yet classical electrodynamic framework for predicting the chiroptical response, in terms of CD spectra, of metal nanostructures across various size regimes. Specifically, we employ the frequency-dependent fluctuating charges and fluctuating dipoles approach (\wfqfmu), which was designed to describe the plasmonic response of metal nanostructures.\cite{giovannini2019classical,giovannini2022we,nicoli2023fully} This method retains a fully atomistic description of the nanomaterial and accounts for both intraband charge transport\cite{giovannini2019classical} and interband polarization within a coupled electrodynamic treatment,\cite{giovannini2022we} by also incorporating quantum tunneling effects.\cite{giovannini2019classical,bonatti2022silico} Crucially, \wfqfmu correctly reproduces the plasmonic features of nanostructures below the quantum size limit,\cite{giovannini2022we} and, at the same time, its favorable scaling enables simulations far beyond the TDDFT limit.\cite{lafiosca2021going,giovannini2025electric,tapani2026morphology} Importantly, \wfqfmu allows for a real-space analysis of the induced densities and current patterns, making it a natural basis for multiscale plasmon-molecule 
approaches.\cite{lafiosca2023qm,sodomaco2026atomistic,morton2010discrete}

Therefore, we exploit the \wfqfmu method to test whether this atomistic electrodynamical model can reproduce and interpret plasmonic spectra with QM accuracy. We perform a systematic analysis that spans the quantum-size regime, where discrete electronic transitions and QM effects are expected to be essential,\cite{karimova2015time} the quantum--plasmonic crossover in atomistically defined gold nanostructures still accessible at the TDDFT level,\cite{toffoli2021circularly,toffoli2021plasmonic} and experimentally studied DNA-origami-directed assemblies of gold nanorods \cite{wang2019reconfigurable} reaching $\sim10^5$ atoms. In the latter, a classical description is expected to be adequate; however, atomic resolution remains valuable for connecting chirality to local morphology and coupling pathways. We compare \wfqfmu results with state-of-the-art TDDFT calculations for atomistically defined nanostructures and experimental CD spectra for large nanoassemblies, finding quantitative agreement in line shapes, sign patterns, and relative intensities. These results show that an atomistic, yet classical, electrodynamic picture can reproduce the optical activity of chiral plasmonic structures when the underlying physics is modeled correctly in terms of intra- and interband transitions. This supports a unified interpretation of plasmonic chirality in terms of geometry-driven collective excitation patterns and, accordingly, questions whether explicit QM treatments are necessary. More broadly, this work establishes the atomistically resolved electrodynamic framework required for predictive modeling of realistic chiral plasmonic architectures and hybrid plasmon--molecule systems, with direct access to the local chiral near fields that govern sensing, catalysis, and chiral light--matter interactions.

\section{Results}

\subsection{From the quantum to the plasmonic regime}

\begin{figure*}[!htbp]
    \centering
    \includegraphics[width=1\linewidth]{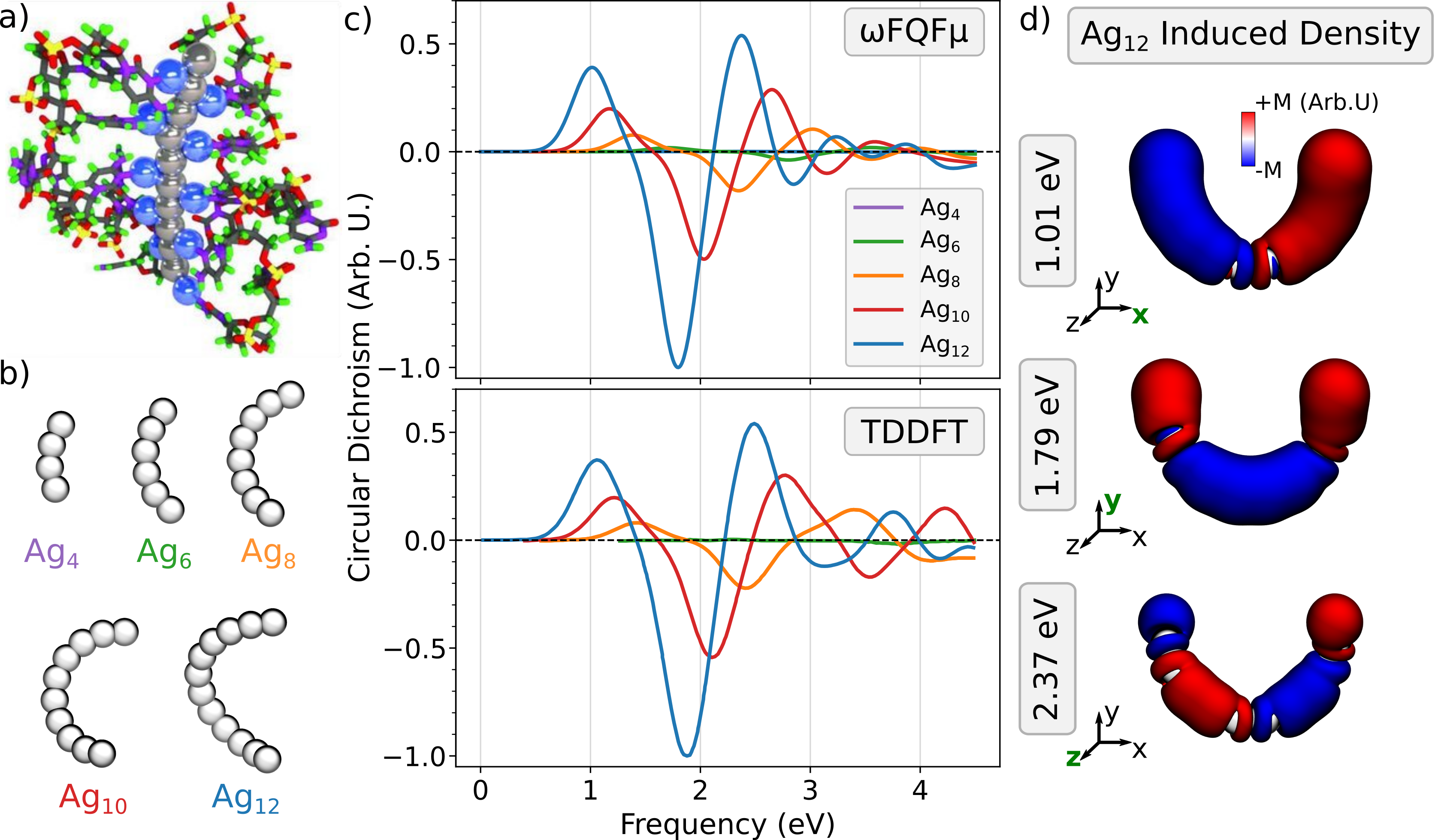}
    \caption{(a) Examples of the Ag:DNA structures in trimer arrangements, adapted from ref \citenum{schultz2013evidence}. (b-c) structures and calculated CD spectra of Ag nanochains as a function of the number of atoms. Reference TDDFT spectra are recovered from Ref. \citenum{karimova2015time}, performed at the SAOP/TZP levels by including relativistic effects \textit{via} ZORA corrections. (d) \wfqfmu imaginary part of induced densities for Ag$_{12}$ at the three main plasmon resonances for the specified field polarizations.}
    \label{fig:ag_chain}
\end{figure*}

We first assess the performance of the atomistic classical model in the quantum-size regime by considering helical silver chains containing between four and twelve atoms. Despite their small size, these structures are experimentally relevant because few-atom Ag clusters can be stabilized by DNA templates (Ag:DNA), where a small number of Ag atoms is hosted within the biomolecular scaffold and gives rise to intense, size-dependent optical signatures (see Fig. \ref{fig:ag_chain}a).\cite{schultz2013evidence} Their optical activity has been characterized through combined absorption and circular dichroism measurements and interpreted with TDDFT calculations for Ag$_n$:DNA complexes with $n=4, 6, 8, 10, 12$. \cite{karimova2015time}
These systems represent a particularly challenging case study for a classical model, because their optical response is expected to be dominated by discrete electronic transitions, as predicted by TDDFT. In particular, TDDFT results have shown that even a slight loss of planarity in small Ag chains is sufficient to induce a pronounced chiroptical activity, which agrees well with the experiments.\cite{karimova2015time}

In Figure~\ref{fig:ag_chain}b,c we consider the Ag$_n$ nanochains with Ag-Ag-Ag bond and Ag-Ag-Ag-Ag dihedral angles fixed to 160 and 10 degrees, respectively.\cite{karimova2015time} Note that since TDDFT results were broadened with a Gaussian line shape,\cite{karimova2015time} we also broadened the \wfqfmu spectra by a Gaussian convolution (see Fig. S1 in the Supplementary Information (SI) for raw data). The TDDFT absorption spectra of the selected Ag helices are characterized by a dominant low-energy peak in the region 1.0--2.5 eV, which redshifts as the number of atoms increases, reflecting the progressive reduction of the HOMO--LUMO gap with chain length (see Figs.~S1 and S2 in the SI). For the smaller clusters ($n<10$), such a low-energy excitation dominates the spectrum, while longer chains ($n>10$) show additional higher-energy features associated with transitions that rapidly become allowed upon loss of linear symmetry. Importantly, the corresponding TDDFT-based CD spectra in Fig.~\ref{fig:ag_chain}c display a $(+,-,+)$ pattern for all helical structures, with the intensity increasing significantly with chain length. Such a sign alternation is a hallmark of chirality at the atomic scale and is directly linked to the helical geometry of the metal backbone. As one can see in Fig.~\ref{fig:ag_chain}c, the atomistic \wfqfmu model reproduces this behavior with impressive accuracy. Both the \wfqfmu absorption spectra (see Fig.~S2 in the SI) and the CD spectra are not only in qualitative but also in quantitative agreement with the TDDFT reference across the entire Ag$_4$--Ag$_{12}$ series, capturing the peak positions and sign pattern. In particular, the systematic redshift of the main absorption band with increasing chain length and the concomitant enhancement of the CD signal are correctly recovered. The model remarkably reproduces the $(+,-,+)$ alternation predicted by TDDFT, and provides an excellent agreement also in the intensities of the CD bands. These results show that, even at the quantum scale, the classical \wfqfmu model reproduces all key chiroptical signatures.

In the TDDFT description, the chiroptical response of these clusters was rationalized in terms of specific discrete molecular orbital (MO) transitions, with the HOMO$\rightarrow$LUMO excitation playing a central role in the low-energy CD band.\cite{karimova2015time} In contrast, \wfqfmu contains no explicit electronic orbitals: it describes the optical response in terms of intraband charge exchange (Drude-like conduction) and interband polarization, both induced by the external field and coupled through the fully atomistic Coulomb interactions (see also Methods). To provide a physical view of the origin of chirality in the studied Ag clusters, we report in Fig.~\ref{fig:ag_chain}d the imaginary part of the induced charge densities at the three most intense CD peaks of the Ag$_{12}$ structure (analogous plots for all other systems are provided in Sec.~S2 of the SI). Note that these peaks are associated with specific field polarizations as graphically depicted in Fig. S3 in the SI, specifically $x$ (1.01 eV), $y$ (1.79 eV), and $z$ (2.37 eV). The three induced densities are characterized by an increasing number of nodes as the excitation frequency increases, and closely resemble the MOs involved in the transitions reported in Ref. \citenum{karimova2015time}. \wfqfmu allows us to disentangle intraband (associated with the Drude conduction mechanism) and interband contributions to the total induced density (see Figs. S4-S8 in the SI). As expected for these rod-like systems,\cite{bonatti2020plasmonic} the induced densities are mainly due to the intraband mechanism,\cite{giovannini2022we} with interband effects mostly affecting the high-energy transitions. More generally, Fig. \ref{fig:ag_chain}d allows us to directly interpret the CD signal from purely geometry-dependent phase relations between induced polarization and the associated magnetic dipole moment, which are parallel at 1.01 and 2.37 eV (and thus the positive CD band), and opposite at 1.79 eV (and thus the negative CD peak). 

The fact that our approach reproduces both the full TDDFT spectra and the underlying spatial patterns indeed suggests that the essential electrodynamical ingredients governing the chiroptical response of these systems are properly accounted for in the purely classical \wfqfmu method.

\begin{figure*}[!htbp]
    \centering
    \includegraphics[width=.95\linewidth]{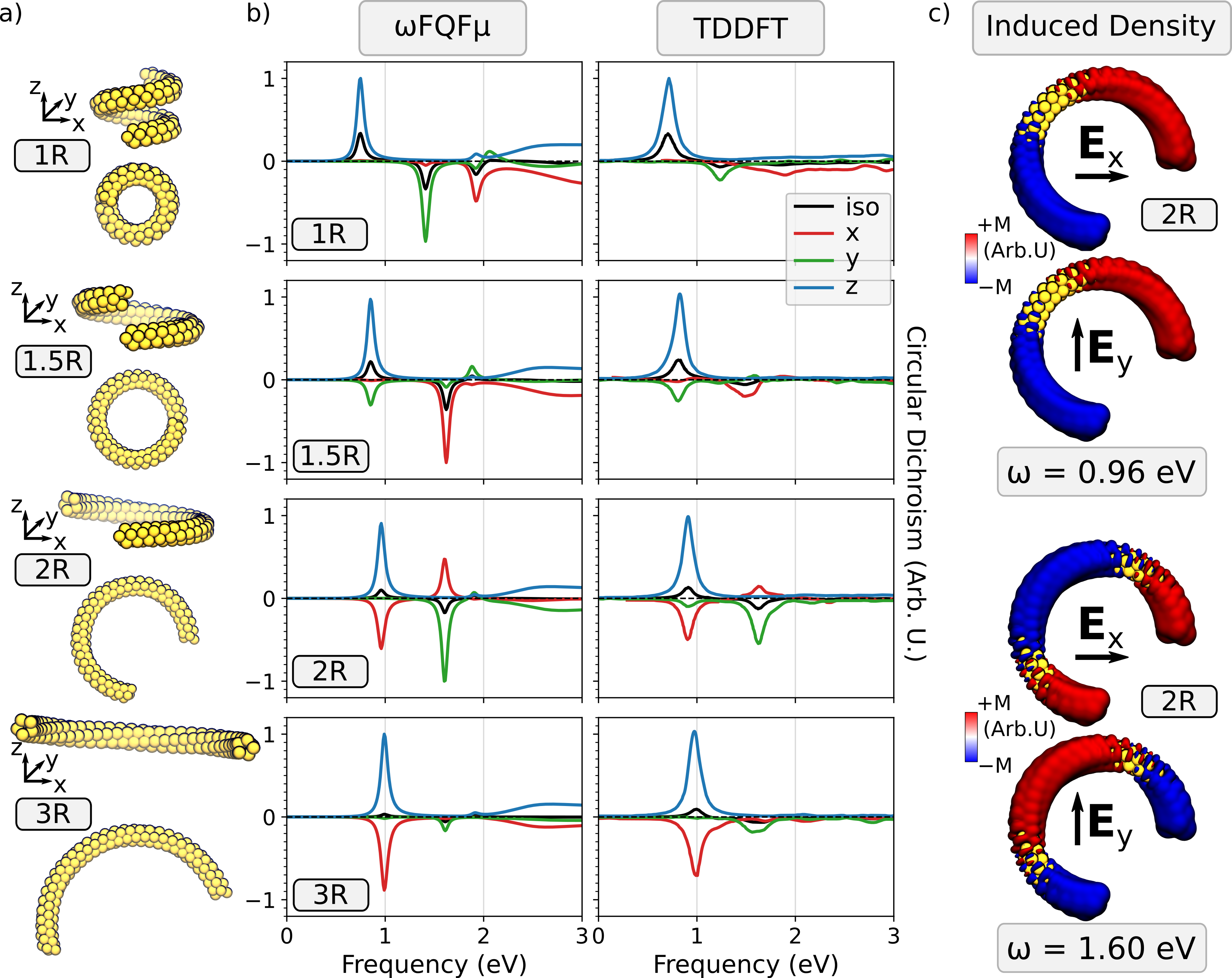}
    \caption{(a) Side and top view of 1R--3R gold helices structures. (b) \wfqfmu (left) and TDDFT (right) CD spectra decomposed in Cartesian components. TDDFT results are recovered from Ref. \citenum{toffoli2021circularly}, and were computed at the polTDDFT/LB94/TZP level, by including relativistic correction \textit{via} ZORA. (c) \wfqfmu imaginary densities (isovalue 0.01 a.u.) computed at the two main plasmon resonances of 2R ($\omega=0.96$ and 1.60 eV) induced by $x$ and $y$-polarized electric field.}
    \label{fig:au_helix}
\end{figure*}

To corroborate this hypothesis, we consider larger systems that bridge the gap between discrete, molecule-like excitations and plasmonic response, while remaining accessible at the TDDFT level. Specifically, we study a chiral gold (5,3) nanotube (length $\sim$ 82~\AA) arranged either as a right-handed helical configuration or as a chiral dimer. Both structures are constructed by following Refs.~\citenum{toffoli2021circularly} and \citenum{toffoli2021plasmonic}, respectively, where they were proposed for enhancing the circular dichroism signals of the linear (5,3) nanotube that displays almost zero chiral response. The right-handed helices are generated by coiling the Au(5,3) nanotube and progressively increasing the helix radius (1R, 1.5R, 2R, and 3R; Fig.~\ref{fig:au_helix}a). The corresponding CD spectra computed with the atomistic electrodynamic model (\wfqfmu) are compared in Fig.~\ref{fig:au_helix}b with the reference TDDFT results from Ref.~\citenum{toffoli2021circularly}. The latter are obtained within an approximate TDDFT formulation (polTDDFT), where, similarly to \wfqfmu, the chiroptical response is reconstructed from the frequency-dependent complex electric response.\cite{baseggio2016extension} At the TDDFT level, all helices display a clear bisignate CD profile in the isotropic spectrum (iso) with a $(+,-)$ sign alternation in the low-energy region ($< 2$~eV), with the intensity of the first positive band decreasing by increasing $R$. The two CD bands are located at the same frequencies as the dominant plasmonic resonances predicted in absorption (see Fig.~S9 in the SI). It is worth pointing out that the bisignate feature in the CD spectrum is typical of helical nanostructures,\cite{bai2016two,hoflich2019resonant} as also reported experimentally.\cite{mark2013hybrid,liu2016chiroptically,esposito2015nanoscale} The same CD line shape, including the correct sign sequence and its evolution across the 1R--3R series, is also reproduced by \wfqfmu. The agreement is not limited to CD: in fact, \wfqfmu also reproduces the absorption spectra, which are dominated by two low-energy plasmonic bands whose peak positions and oscillator strengths evolve systematically with the helix radius (see Fig.~S9 in the SI), with the lowest-energy band increasing in intensity upon increasing the coiling radius.

As stated above, both polTDDFT and \wfqfmu evaluate CD within a complex-response framework. Therefore, we can extend the comparison between the methods by separately considering the three diagonal components of the rotational strength tensor (Fig.~\ref{fig:au_helix}b; see also Fig.~S10 in the SI). This way, we provide a direct assessment of how the individual Cartesian contributions evolve with coiling and how they combine into the final $(+,-)$ bisignate spectrum. Such Cartesian decomposition indeed clarifies the systematic evolution across the 1R--3R series. In the most tightly coiled helix (1R), the low-energy positive CD band at about 0.75 eV is dominated by the $z$-component, parallel to the helix axis, whereas the negative contribution at higher energy ($\sim$ 1.4 eV) is mainly due to the transverse $y$ component, with the $x$ component contributing to the peaks arising at higher energies. Along the series, the longitudinal $z$-component contributing to the low-energy positive band is consistently reproduced with the same relative intensity. However, a negative transverse contribution ($y$ for 1.5R, and $x$ for 2R and 3R) substantially increases by increasing the radius, yielding an overall decrease of the positive band in the isotropic average. The diverse predominant transverse contribution directly follows from the helix geometry and is also reported by the first absorption band, which exhibits an analogous change in its dominant Cartesian character (see Fig. S10 in the SI). The apparent reduction of the averaged rotatory strength for larger radii thus does not imply a weakening of the longitudinal plasmonic contribution, but rather reflects the growing weight of transverse terms opposing it. This trend is reproduced by \wfqfmu at the same level of detail as TDDFT,\cite{toffoli2021circularly} including the components dominating each CD peak (Fig.~\ref{fig:au_helix}b). Only minor discrepancies in relative intensities are observed, which are due to the differences in the effective band widths reproduced by the two approaches.

The bisignate $(+,-)$ feature of the CD spectrum can be rationalized by viewing each plasmonic resonance as a collective current oscillation constrained by the helical geometry. Such current is related to the induced densities that underlie each CD signal, providing a real-space analysis. The longitudinal mode (along the helix axis, $z$) induces an electric dipole aligned with $z$ (see Figs.~S11-S14 in the SI for induced densities) and a current flowing along the helix. Due to the right-handed current path, the associated magnetic dipole is aligned with the electric one. The corresponding contribution to the rotational strength is therefore positive throughout the considered series. Let us now consider transverse excitations ($x$ and $y$ directions). As an illustrative case, we focus on the 2R helix, for which the two $(+,-)$ CD bands involve both $x$- and $y$-induced transverse responses. The \wfqfmu induced densities are shown in Fig.~\ref{fig:au_helix}c. At both resonances, intra- and interband transitions contribute significantly to the total response (see Fig.~S13 in the SI), demonstrating that, in systems more complex than Ag nanochains, chiral nanoplasmonics cannot be described correctly unless both mechanisms are treated in a physically consistent way. In this respect, a distinctive feature of \wfqfmu is that intraband and interband effects are encoded into separate microscopic variables (charges and dipoles, respectively; see Methods), which allows their contributions to be disentangled, providing direct physical insight into the origin of the CD signal.

For the low-energy resonance ($\omega = 0.96$~eV), the $x$- and $y$-polarized induced densities are in phase (same overall sign pattern across the structure). Consequently, the transverse contributions combine constructively in the CD response, consistent with the component-resolved spectra in Fig.~\ref{fig:au_helix}b at both \wfqfmu and TDDFT levels.\cite{toffoli2021circularly} In contrast, for the high-energy mode ($\omega = 1.60$~eV), the $x$ and $y$ responses are phase-opposed, leading to a sign inversion of the corresponding transverse contribution to the overall CD signal. In particular, the charge flows across the helix cross-section, producing an electric dipole perpendicular to the helix axis, while the induced current is split into two branches along the turns, which effectively contribute with opposite handedness with respect to the transverse dipole (see Fig.~\ref{fig:au_helix}c.). As a result, the magnetic dipole tends to oppose the electric one, and the associated rotational-strength contribution becomes positive or negative depending on the sign of the electric dipole moment. Remarkably, the same interpretation was achieved based on TDDFT calculations\cite{toffoli2021circularly} (see Fig. S13 in the SI for a graphical comparison between \wfqfmu and TDDFT induced densities). The $(+,-)$ CD profile thus reflects this competition between a positive longitudinal term and negative transverse terms, whose relative weight changes with the helix radius and is directly quantified by the tensor-component comparison (see Fig.~\ref{fig:au_helix}b). 

\begin{figure*}[!htbp]
    \centering
    \includegraphics[width=1\linewidth]{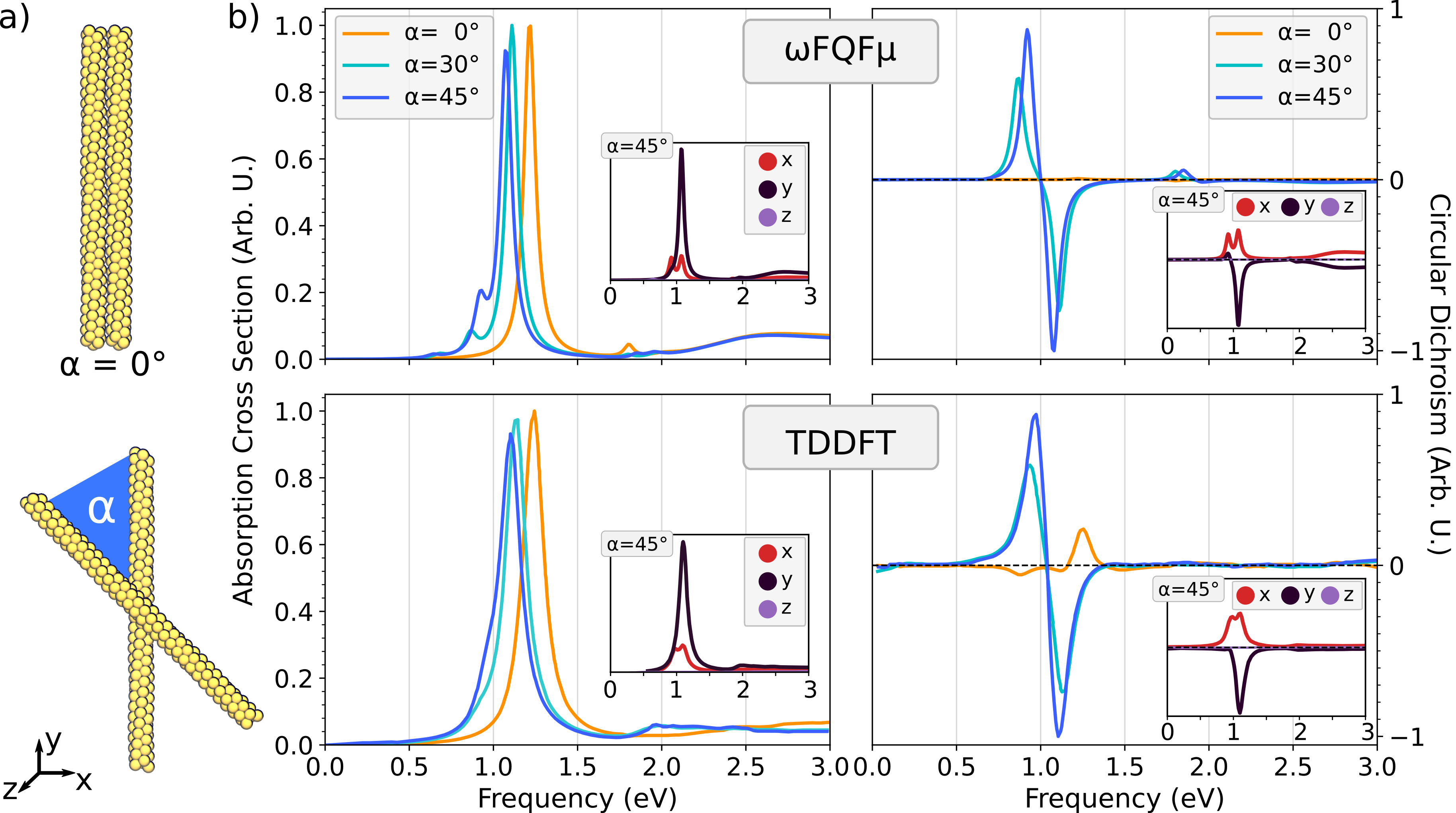}
    \caption{(a) Side and top view of parallel (top) and tilted gold assembled nanotubes. (b) \wfqfmu (top) and TDDFT (bottom) absorption (left) and CD (right) spectra of assembled nanotubes as a function of the tilting angle. TDDFT results are recovered from Ref. \citenum{toffoli2021plasmonic}, and were computed at the polTDDFT/LB94/TZP level, by including relativistic corrections \textit{via} ZORA. The Cartesian decomposition of the absorption and CD signals for $\alpha = 45^{\circ}$ is given as insets.}
    \label{fig:au_X}
\end{figure*}

We next consider dimer assemblies of chiral Au(5,3) nanowires, which are of particular interest because strong plasmonic CD signals are typically observed in assembled nanostructures where near-field coupling amplifies chiroptical activity. This is also one of the most widely used routes to generate chirality in nanoplasmonics.\cite{kuzyk2012dna,zhou2017dna,shen2013three,ma2013attomolar,lan2016self,kneer2018circular} It is therefore crucial that a predictive model for chiral nanoplasmonics is able to describe not only isolated nanostructures, but also coupled assemblies. In this work, we follow Ref. \citenum{toffoli2021plasmonic} by placing two chiral Au(5,3) nanotubes at an interwire distance of 2.88~{\AA} (bulk Au--Au distance). In particular, we investigate three relative orientations obtained by rotating one nanowire around the common $C_2$ ($z$) axis by $\alpha=0^\circ$ (parallel configuration, see Fig. \ref{fig:au_X}a top), $30^\circ$, and $45^\circ$ (see Fig.~\ref{fig:au_X}a bottom). To show the reliability of \wfqfmu also for chiral assemblies, we compare the computed absorption and CD spectra with the polTDDFT reference data from Ref. \citenum{toffoli2021plasmonic}.

The absorption spectra of the three assemblies computed at both levels are graphically depicted in Fig. \ref{fig:au_X}b left panel. \wfqfmu spectra are characterized by a main absorption peak which is located at about 1.25 eV for the parallel configuration. Upon rotation ($\alpha=30^\circ$ and $45^\circ$), the main peak shifts to lower energies, reflecting the reduced destabilizing interaction when the induced dipoles are no longer parallel. The spectra of the tilted configurations are also characterized by an additional shoulder at lower energy ($< 1$ eV), while at higher energies ($> 1.5$ eV) interband transitions dominate the spectrum, as expected. The agreement with the reference TDDFT absorption spectra\cite{toffoli2021plasmonic} is once again excellent, with \wfqfmu correctly reproducing the overall shifts upon rotating the assemblies. The only minor discrepancy concerns the band widths, which, as in the case of Au helices, are predicted to be larger at the TDDFT level. As a consequence, the shoulders predicted by \wfqfmu are partly absorbed within the main plasmonic band. 

The CD response is significantly more sensitive to the relative orientation (see Fig. \ref{fig:au_X}b right panel). For $\alpha=0^\circ$, \wfqfmu predicts almost zero CD signal, consistent with the fact that the parallel dimer arrangement is an achiral morphology and the chiral monomers show almost zero CD signals. Increasing the rotation angle leads to a strong CD signal. At both $\alpha=30^\circ$ and $\alpha=45^\circ$, the spectrum is characterized by a bisignate $(+,-)$ profile with two intense peaks of opposite sign separated by a small energy splitting. The emergence of the CD signal is due to the transverse channels with respect to the main assembly axis (see Fig. \ref{fig:au_X}b insets for $\alpha = 45^\circ$ component-resolved spectra). In fact, both the $x$ and $y$ dipolar components contribute to the overall sign alternation, and their balance controls the sign and magnitude of each CD band (see also Fig.~S15 in the SI). 

Similarly to the helix case, the real-space induced densities provide a complementary microscopic analysis. For the rotated dimers with $\alpha=45^\circ$ (see Fig. S16 in the SI), the induced density patterns at the energies corresponding to the CD maximum and minimum display clear dipolar charge separation on each wire, confirming the plasmonic character of both features. The relative arrangement of the induced dipoles on the two wires changes between the two main CD resonances. The first one corresponds to an anti-aligned dipolar configuration between the two nanowires, whereas the other corresponds to an aligned configuration. This switch in the relative phase/orientation of the induced currents modifies the effective electric-magnetic coupling within the assembly, and results in opposite CD signs for the two bands. Remarkably, the \wfqfmu induced-density plots (see Fig.~S16 in the SI) almost perfectly reproduce the TDDFT results, demonstrating again that the model reproduces the reference \textit{ab initio} results,\cite{toffoli2021plasmonic} and suggesting once again that a purely classical model can describe the correct physical mechanisms behind the phenomenon. 


\subsection{Towards Realistic Systems}

\begin{figure*}[!htbp]
    \centering
    \includegraphics[width=.9\linewidth]{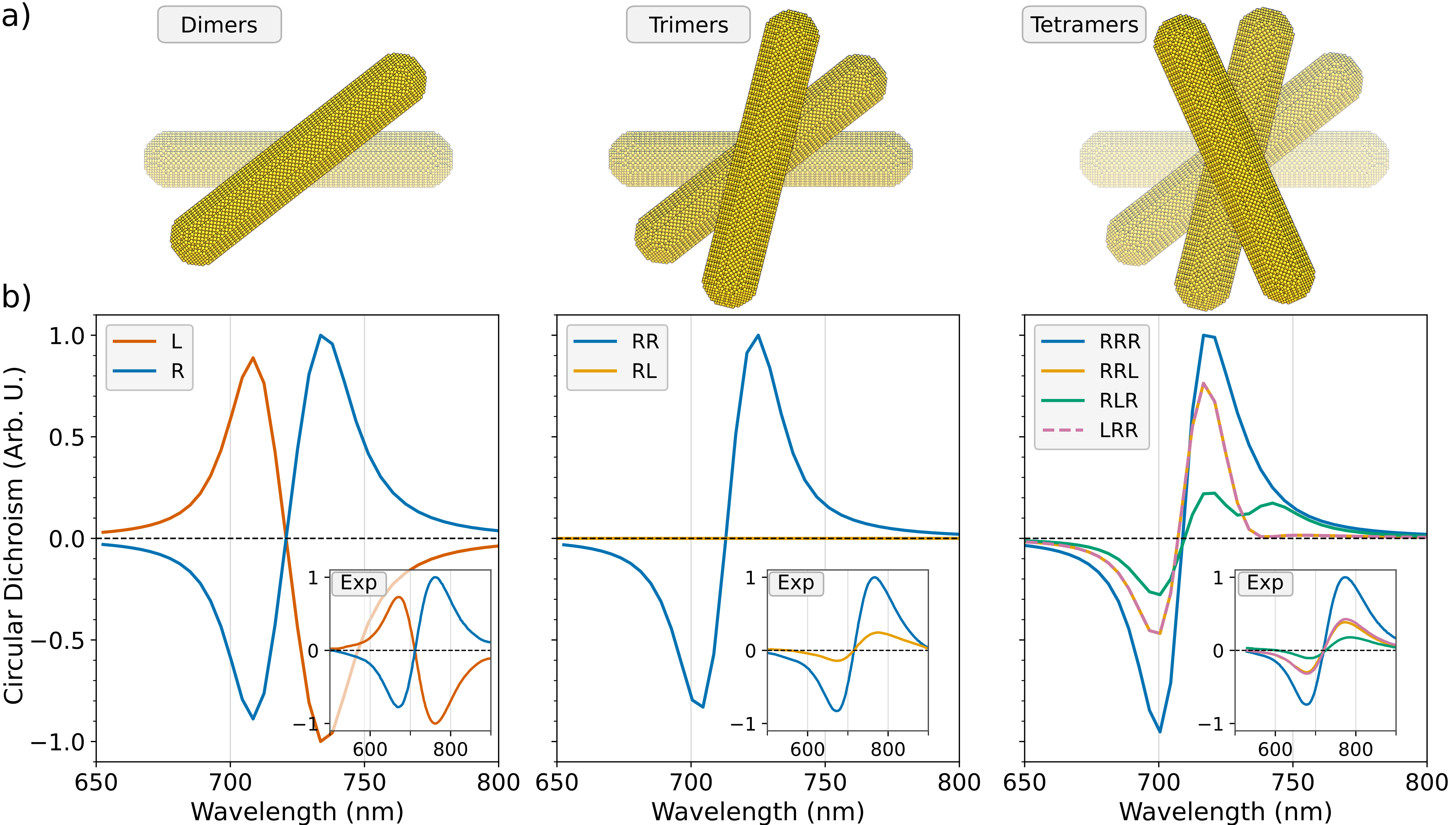}
    \caption{Structures (a) and \wfqfmu CD spectra (b) of chiral DNA-origami-templated gold nanorod assemblies in dimer (left), trimer (middle), and tetramer (right) configurations. The experimental spectra from Ref. \citenum{wang2019reconfigurable} are given as insets.}
    \label{fig:rod_spiral}
\end{figure*}

As a final step towards experimentally realistic systems, we consider DNA-origami-templated gold nanorod assemblies that form reconfigurable plasmonic enantiomers and diastereomers.\cite{wang2019reconfigurable} Remarkably, DNA-origami-based technology represents one of the most widely adopted approaches for fabricating chiral plasmonic assemblies with precise geometrical control.\cite{kuzyk2012dna,zhou2017dna,shen2013three,ma2013attomolar,lan2016self,kneer2018circular} The structures considered in this work are experimentally constructed by considering individual nanorods ($12$~nm x $42$nm) arranged at a separation of about 22 nm to create chiral crossed geometries (rotation angle of 38$^\circ$) whose handedness can be reversibly switched.\cite{wang2019reconfigurable} Multiple such chiral centers can then be hierarchically combined to generate trimers and tetramers with prescribed handedness sequences (R, L; RR, RL; RRR, RRL, RLR, LRR, etc.), thus enabling the controlled realization of both enantiomeric and diastereomeric assemblies.\cite{wang2019reconfigurable} To provide a direct comparison with the experiment, we first construct a model nanorod with dimensions chosen to reproduce the experimental plasmon resonance in the 700--750~nm spectral range for the individual nanostructure (see Fig. S17 in the SI). The resulting nanorod is composed of 22308 atoms (4.5 nm x 25.2 nm). We then construct crossed dimers (R and L, 44616 atoms), trimers (RR and RL, 66924 atoms), and tetramers (RRR, RRL, RLR, LRR, 89232 atoms) by imposing an inter-rod distance of 11 nm and a rotation angle of $38^\circ$ (see Fig. \ref{fig:rod_spiral}a and Fig. S18a in the SI). It is worth remarking that at these length scales, the (chiro)optical response can also be described by exploiting continuum classical electrodynamics methods,\cite{wang2019reconfigurable} such as the boundary element method.\cite{de2002retarded} Here, we demonstrate that \wfqfmu can be used to study experimentally relevant nanostructures while retaining a fully atomistic description. Our method thus paves the way for studying experimentally accessible architectures where atomistic features (e.g., local morphology and contact regions) can affect the (chiro)optical response.

The calculated CD spectra for the full series of assemblies are reported in Fig.~\ref{fig:rod_spiral}b. As expected, the single chiral center (R and L) shows a bisignate profile with opposite signs since the two structures are enantiomers. Also in this case, we decompose the absorption and chiroptical signal into Cartesian components (see Fig. S18b in the SI). While the absorption is dominated by the interplay between the transverse ($x$ and $y$) components, the CD arises primarily from the $y$ component. The induced densities (see Fig. S19 in the SI) provide a graphical visualization for the origin of the opposite CD signal for the two bands: in the high-energy band, the induced densities on the two rods oscillate in phase and generate a magnetic dipole in the same direction of the electric dipole, while in the low-energy band they oscillate with opposite phases, generating an opposite magnetic dipole.
The homochiral trimer RR shows the same $(-,+)$ sign alternation reported for the R dimer. However, in this case, the component analysis (see Fig. S18b in the SI) reveals that both $x$ and $y$ channels contribute to both CD bands. Remarkably, similarly to the previously considered chiral assemblies (see Fig. \ref{fig:au_X}b), the resulting CD profile is a delicate balance between the CD signals arising from both transverse polarizations. The CD signal of the diastereomeric trimer RL instead shows an identically zero CD spectrum, since the structure is not chiral. The non-zero experimental CD reported for this structure\cite{wang2019reconfigurable} likely arises from an excess of R dimers within the trimer configurations.

Moving to tetramers, all assemblies preserve the same $(-,+)$ profile reported by the homochiral RR trimer and the R dimer. In all cases, the Cartesian decomposition reported in Fig.~S18b in the SI shows that the overall bisignate CD signal is due to both transverse components, which combine to provide the isotropic CD spectrum. In particular, for the homochiral RRR assembly, the sign alternation is mainly due to the $x$ polarization, while in all other cases, $x$ and $y$ responses partially counterbalance to provide the final bisignate signals. Remarkably, for all other tetramer assemblies (RRL, RLR, and LRR), the chiral response is lower in intensity as compared to the homochiral RRR, reflecting the partial compensation introduced by the L center. In particular, for RLR, where opposite-handed centers alternate, the cancellation becomes substantially more effective, and the CD is strongly suppressed over the whole spectral range. In this configuration, the $x$ and $y$ contributions to the rotational strength are nearly balanced in magnitude but opposite in sign, yielding a small isotropic average. These trends are confirmed in real space by the induced densities (see Figs.~S19-S23 in the SI): compared to RRR, the mixed assemblies display induced densities that are characterized by phase-opposed current pathways between neighboring centers as induced by perpendicular fields, which weaken the net effective electric--magnetic dipole coupling of the assembly and reduce the overall CD.

Importantly, all the calculated spectra are in very good agreement with the experimental data\cite{wang2019reconfigurable} reported in the insets of Fig.~\ref{fig:rod_spiral}b. In particular, the model correctly reproduces the experimental sign alternation and the relative intensities among the CD peaks and within the same series. A slight deviation is associated with linewidths, as the experimental bands are generally broader. This is expected for DNA-origami nanostructures, where ensemble effects such as structural heterogeneity, gap dispersion, and angular fluctuations inherent to DNA-origami systems, which are not explicitly included in the present single geometry calculations, provide additional inhomogeneous broadening mechanisms. This can be appreciated by comparing \wfqfmu and experimental absorption spectra of the monomer (see Fig. S17 in the SI). Within these limits, the agreement shows that \wfqfmu captures the dominant mechanism controlling optical activity in these large assemblies in terms of near-field coupling between plasmonic building blocks.

\section{Discussion}

We have shown that the fully atomistic classical electrodynamic \wfqfmu model can compute chiroptical response in metallic nanostructures across length scales, from few-atom clusters to experimentally realistic plasmonic assemblies containing up to about a hundred thousand atoms. Ranging from few-atom silver nanochains in the quantum-size regime, to gold nanotube helices at the molecule-plasmon crossover, coupled dimers, and DNA-origami-templated nanorod assemblies, the model quantitatively reproduces reference TDDFT calculations and available experimental spectra, capturing peak positions, sign alternations, and relative intensity trends.

Our results clarify the physical origin of plasmonic chirality in these systems. In particular, they show that an explicit QM treatment is not required to reproduce the dominant chiroptical response, even at the atomic scale. Indeed, while quantum approaches remain crucial for determining the electronic structure and related effects, such as hot-carrier energy distribution, the present analysis indicates that the dominant CD signatures are governed by geometry-driven collective currents and their electromagnetic coupling. As such, chiroptical properties can be described by a classical model, as long as this model correctly retains the atomic-scale morphology and the main physical ingredients describing the plasmonic response. A key aspect is the physically consistent treatment of intra- and interband effects, which is essential to reproduce complex chiroptical signals and to disentangle their microscopic origin. At the same time, \wfqfmu can describe nanostructures of sizes far beyond those treatable at the \textit{ab initio} level, while preserving the local atomic structure that can determine optical activity through defects, contact regions, and ultrasmall gaps. This makes the method suitable not only for rationalizing experiments but also as a predictive tool for designing chiral plasmonic architectures where both global optical response and local near-field chirality must be controlled. 


By demonstrating that plasmonic chirality can be accurately described within an atomistic, yet classical, electrodynamic approach, this work paves the way for systematic studies of chiral optical activity at experimental length scales while retaining access to its atomistic origin. This is particularly relevant for hybrid plasmon--molecule systems, where local chiral near fields and atomistic motifs can strongly influence the chiroptical molecular response in sensing and catalytic applications.\cite{abdali2008surface,bainova2023plasmon,both2022nanophotonic,serrano2026chiral} 
By providing the missing foundation to connect local structural motifs to both chiroptical response and local chiral near fields, this work opens the way to the atomistically informed, rational design of chiral plasmonic nanostructures and hybrid plasmon--molecule systems optimized for targeted applications, from enantioselective sensing and asymmetric catalysis to chiral light--matter interactions in both the weak- and strong-coupling regimes, reserving the quantum description to the molecular degrees of freedom that genuinely demand it.


\section{Methods}

In this paper, we exploit the atomistic frequency-dependent fluctuating charges and dipoles (\wfqfmu) model.\cite{giovannini2019classical,giovannini2022we} In this framework, each metal atom $i$ is assigned (i) a complex charge $q_i(\omega)$, which models the intraband (Drude-like) conduction mechanism,\cite{giovannini2019classical} and (ii) a complex induced dipole moment $\boldsymbol{\mu}_i(\omega)$, which accounts for interband transitions through a frequency-dependent atomic polarizability.\cite{giovannini2022we} For a monochromatic external field oscillating at frequency $\omega$, the induced charges and dipoles are obtained in the quasi-static approximation by solving the coupled linear problem defined by:\cite{giovannini2022we}
\begin{align}
\frac{dq}{dt} = -i\omega\, q_i(\omega) &= \frac{2n_0\tau}{1-i\omega\tau}\sum_{j}\!\left[1-f(l_{ij})\right]\frac{\mathcal{A}_{i}}{l_{ij}}\left(\phi_j-\phi_i\right),
\label{eq:methods_wfq_short}\\[4pt]
\boldsymbol{\mu}_i(\omega) &= \alpha_i^{\omega}\Big(\mathbf{E}^{ext}_i+\mathbf{E}^{q}_i+\mathbf{E}^{\mu}_i\Big),
\label{eq:methods_wfmu_short}
\end{align}
where $n_0$ is the electron density, $\tau$ is the scattering time, $\mathcal{A}_{i}$ is the effective area through which the charge flows, $l_{ij}$ is the distance between atoms $i$ and $j$, and $f(l_{ij})$ is a Fermi-like function mimicking quantum tunneling. $\phi_i$ is the electrochemical potential acting on atom $i$, including Coulomb interactions (charge--charge, charge--dipole, dipole--dipole\cite{giovannini2019fqfmu}) and coupling to the external field.\cite{giovannini2022we} $\mathbf{E}^{q}_i$ and $\mathbf{E}^{\mu}_i$ are the local fields generated by the induced charges and dipoles, respectively, while $\alpha_i^{\omega}$ is the complex atomic polarizability used to mimic interband contributions, extracted from the experimental permittivity.\cite{giovannini2022we,nicoli2023fully}  Solving Eqs.~\eqref{eq:methods_wfq_short}--\eqref{eq:methods_wfmu_short} provides the charges and dipoles at frequency $\omega$, from which the absorption, scattering, and extinction cross sections can be computed.\cite{giovannini2026plasmonx} 

In this work, we extend \wfqfmu to the calculation of electronic circular dichroism (CD) signals,\cite{barron2004molecular} which is the difference in absorption of left and right circularly polarized light, and is expressed in terms of molar extinction coefficients $\Delta \epsilon (\omega)$:\cite{jurinovich2018exat,makkonen2021real}
\begin{equation}
    \Delta \epsilon(\omega) = \frac{16 \pi N_A}{3 \ln(10) 10^3} \frac{2\pi}{\hslash c} \omega R(\omega)
\end{equation}
where $c$ is the speed of light, $\hslash$ is the reduced Planck constant, $N_A$ is Avogadro’s constant. The CD signal is proportional to $R(\omega)$, which is the rotatory strength density in cgs units,\cite{makkonen2021real} that can be written from the mixed electric dipole-magnetic dipole polarizability tensor, also known as the optical rotatory response tensor, $\beta_{kk}$.\cite{varsano2009towards,makkonen2021real,goings2016atomic} By assuming that the intrinsic magnetic dipole moment is negligible, which is the case for common metal nanostructures, the optical rotatory response tensor can be in turn expressed in terms of the magnetic dipole moment $\mathbf{m}^{(k)}$ induced by an electric field polarized along the $k$ direction (see Section S1 in the SI) as:\cite{varsano2009towards,makkonen2021real,goings2016atomic}
\begin{equation}
    R(\omega) = \frac{\omega}{\pi c} \text{Im} \left[ \sum_k \beta_{kk} (\omega)\right] = \frac{1}{\pi E_0} \text{Re} \left[ \sum_k m^{(k)}_{k} (\omega)\right]
    \label{eq:rotational_starting}
\end{equation} 
where $E_0$ is the intensity of the external field and $k$ indicates Cartesian coordinates (k $\in \{x, y, z\}$). 

In \wfqfmu, the total induced magnetic dipole moment induced by a field polarized along the $k$ axis in the frequency domain can be decomposed in terms of charge $\mathbf{m}^{(k)}_q(\omega)$ and dipole $\mathbf{m}^{(k)}_\mu(\omega)$ contributions as:
\begin{equation}
\mathbf{m}^{(k)}(\omega) = \mathbf{m}^{(k)}_q(\omega) + \mathbf{m}^{(k)}_\mu(\omega)
\label{eq:mag_dipole_decomposition}
\end{equation}
By assuming that the induced current flows linearly between atoms $j$ and $i$, the charge contribution to the induced magnetic moment ($\mathbf{m}^{k}_q$) in the frequency domain reads:
\begin{equation}
\mathbf{m}_q^{(k)}(\omega) = \sum_{i>j} I^{(k)}_{ji}(\omega) \left( \mathbf{r}_i \times \mathbf{r}_{j} \right) 
\end{equation}
where $\mathbf{r}_{i}$ is the position of atom $i$, and $I^{k}_{ji}$ is the electric current moving between atoms $i$ and $j$. Its expression can be directly recovered from the charge equation of motion in Eq. \ref{eq:methods_wfmu_short} (see also Sec. S1 in the SI).
The dipole contribution to the induced magnetic moment ($\mathbf{m}^{k}_\mu$) can be instead calculated directly from the classical definition of the magnetic moment in terms of the polarization vector, which gives (see also Sec. S1 in the SI):
\begin{equation}
  \mathbf{m}_\mu^{(k)}(\omega) = -\frac{i\omega}{2}\sum_{i} \mathbf{r}_i \times \bm{\mu}^{(k)}_i(\omega)
\end{equation}
Within our approach, while the single components of the magnetic dipole moment depend on the choice of the origin, the trace of the magnetic tensor in Eq. \ref{eq:mag_dipole_decomposition} is correctly invariant upon a shift of the origin because the electric polarizability tensor is symmetric (see Sec. S1 in the SI for the complete demonstration). 
Finally, we remark that, in \wfqfmu, charges and dipoles are associated with different physical mechanisms, representing intra- and interband transitions, respectively.\cite{giovannini2022we,tapani2026morphology} This allows us to dissect their respective effects on the overall CD signals, providing a direct way for interpreting the origin of optical activity. 

The calculation of CD spectra is implemented in a development version of the open-source code \texttt{plasmonX}.\cite{giovannini2026plasmonx} All calculations are performed by solving \wfqfmu equations on a 0.01~eV energy grid, using \wfqfmu parameters for Ag and Au atoms from Ref. \citenum{giovannini2022we}.

\section*{Acknowledgements}
TG is grateful to Stefano Corni (University of Padova, Italy) for useful discussions. TG acknowledges financial support from the University of Rome Tor Vergata ``Ricerca Scientistica di Ateneo 2024'' INNOVATIONS. VS acknowledges financial support from the European Union – Next Generation EU in the framework of the PRIN 2022 PNRR project POSEIDON – Code P2022J9C3R. Networking within the COST Action CA21101 “Confined molecular systems: from a new generation of materials to the stars” (COSY) supported by COST (European Cooperation in Science and Technology) and within the International Research Network IRN nanoalloys is also acknowledged. N.M. acknowledges support from the Swedish Research Council (Grant No. 2025-04734), the Knut and Alice Wallenberg Foundation (Grant No. 2023.0089), and the European Research Council (ERC Starting Grant No. 101116253 “MagneticTWIST”).

\section*{Supplementary Information}
Additional methodological details on \wfqfmu and on the complex-response evaluation of CD spectra. Raw spectra, Cartesian decompositions, and polarization-resolved induced densities for Ag$_{4-12}$ nanochains, Au(5,3) helices (1R--3R), Au(5,3) nanotube dimers ($\alpha=0^\circ,30^\circ,45^\circ$), and the DNA-origami Au nanorod assemblies (dimers, trimers, and tetramers).

\bibliography{biblio}

\end{document}